\documentclass{article}
\usepackage{spconf,amsmath,graphicx}
\usepackage{multirow}
\usepackage{subcaption}

\title{Universal Medical Imaging Model for Domain Generalization with Data Privacy}
\name{Ahmed Radwan, Islam Osman, Mohamed S. Shehata}
\address{University of British Columbia\\
Department of Computer Science\\
3333 University Way, Kelowna, BC
}
\begin{document}
%
\maketitle
\begin{abstract}
Achieving domain generalization in medical imaging poses a significant challenge, primarily due to the limited availability of publicly labeled datasets in this domain. This limitation arises from concerns related to data privacy and the necessity for medical expertise to accurately label the data. In this paper, we propose a federated learning approach to transfer knowledge from multiple local models to a global model, eliminating the need for direct access to the local datasets used to train each model. The primary objective is to train a global model capable of performing a wide variety of medical imaging tasks. This is done while ensuring the confidentiality of the private datasets utilized during the training of these models. To validate the effectiveness of our approach, extensive experiments were conducted on eight datasets, each corresponding to a different medical imaging application. The client's data distribution in our experiments varies significantly as they originate from diverse domains. Despite this variation, we demonstrate a statistically significant improvement over a state-of-the-art baseline utilizing masked image modeling over a diverse pre-training dataset that spans different body parts and scanning types. This improvement is achieved by curating information learned from clients without accessing any labeled dataset on the server.
\end{abstract}

\begin{figure*}[!ht]
\centering
    \includegraphics[width=\textwidth]{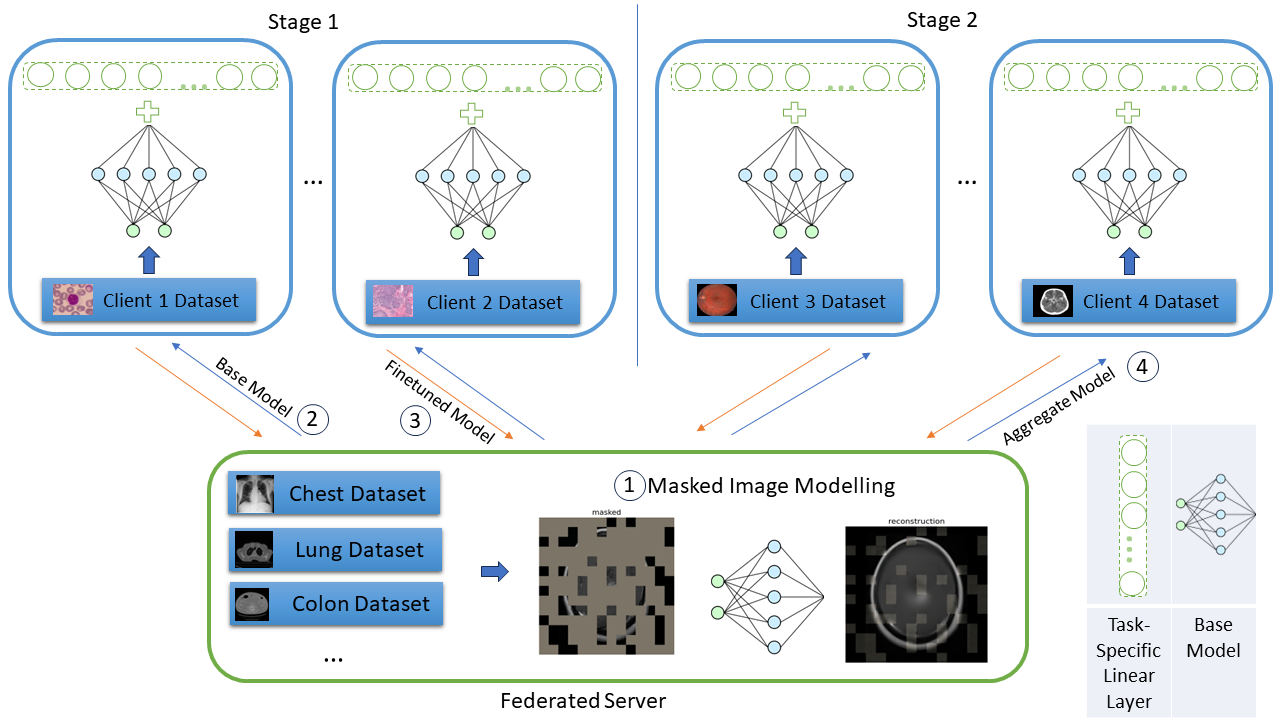}
    \caption{The figure illustrates our Stage-wise federated learning pipeline. 1) The server initializes its base model via self-supervised MIM on data spanning different parts of the human body obtained via different scanning techniques to maintain domain diversity. 2) Clients query the server with the number of classes in their classification problem, and the server sends its base model followed by a randomly initialized linear layer. 3) The client fine-tunes the base model and sends the fine-tuned model back to the server. After receiving a pre-determined number of clients' fine-tuned models, the server concludes the stage, aggregates the weights, and updates the base model as explained in the methodology. 4) The server then sends the updated model to the clients in the next stage.}
    \label{fig:fed}
\end{figure*}
\section{Introduction}
\label{sec:intro}

In the past decade, the field of machine learning (ML), particularly deep learning (DL), has witnessed significant advancements, paving the way for successful applications across various domains. Healthcare stands out as a field with immense potential for transformation due to these DL breakthroughs. The application of DL in healthcare not only holds the promise of directly impacting human life but also offers the potential to alleviate the burden on medical professionals and reduce diagnostic errors. Despite the remarkable achievements of DL techniques in healthcare, three primary challenges demand attention.

Firstly, the success of DL often relies on scaling models to incorporate billions of parameters, necessitating the use of massive datasets. While datasets are typically present on the same machine as the model during training, the practical scenario involves the collection of datasets from various locations. These datasets are then aggregated and transmitted to a central server for model training, introducing communication overhead, particularly when dealing with massive datasets.

Secondly, the healthcare domain presents an added layer of complexity due to the stringent protection of personal health information (PHI) by regulatory laws such as HIPAA and PIPEDA. Safeguarding patient data privacy is a critical consideration in the development and deployment of machine learning models in healthcare.

Finally, ML applications in healthcare are diverse, and they usually involve different types of image scanning techniques. This introduces huge domain shifts and requires models specializing in each domain to achieve good performance, which demands a lot of resources. Moreover, isolated models cannot leverage the commonalities that may exist across these tasks, which can be used to improve the performance across all the tasks. Additionally, it is common to have models that work well on a specific dataset but fail to generalize well when a significant domain shift is present. This can particularly raise concerns regarding the trustworthiness of the model, especially when the model decisions can be critical to decision-making regarding patients' health.

In this paper, we propose a novel method dubbed MedUniverse to address the aforementioned challenges. Our contributions can be summarized as follows:
\begin{itemize}
    \item We collect diverse datasets that span most medical image classification applications to train and test the domain generalization capabilities of our framework.
    \item We utilize a federated learning framework for constructing our MedUniverse model, where each client can protect the privacy of their datasets while allowing the server's model to learn from the finetuned client weights
    \item We achieve strong performance on a variety of tasks where a significant domain shift exists, moreover we show improvements across all tasks even when the domains strongly vary.
\end{itemize}

\section{Background \& Related Work}
\label{sec:related}
\subsection{Domain Generalization}
Standard machine learning approaches can demonstrate formidable performance on specific tasks by splitting the dataset and training on a subset. This assumes that the training and testing data are Independent and identically distributed (i.i.d). However, this is rarely the case with any real-life problem since the testing data distribution can shift over time due to many considerations. In the medical domain, this can prevail in the form of hardware degradation, changes in the scanner hardware, or a different distribution of patients. The domain shift issue is even more prevalent when we completely switch to a different domain, and it usually leads to a huge drop in performance, as shown in work like \cite{walsh2022automated}.
The task of domain generalization can be considered as a form of transfer learning, where we would like to have a model that can generalize to distributions unseen at the test time. The term was first introduced in \cite{muandet2013domain}, where a Domain-Invariant Component Analysis approach was introduced to learn a transformation of the data that minimizes the dissimilarities across the different domains while maintaining a valid input-output mapping through the model. 

\subsection{Masked Image Modelling}
The common scheme in domain generalization is learning domain invariant representations that can be generalized to multiple domains. The idea of masked modeling was first applied in the field of Natural Language Processing (NLP) \cite{devlin-etal-2019-bert} to train transformers in a self-supervised manner. Inspired by this success, \cite{he2022masked}  applies the same technique in computer vision. In Masked Image Modelling (MIM), the image is divided into a set of patches, where a subset is randomly masked out, and the network is trained to reconstruct the hidden patches. The paper first analyzes the key differences between the visual and linguistic input, then proposes an encoder-decoder architecture with a bigger masking ratio for the input image patches to solve this task. The MIM can serve as a pretext task, where the pre-trained encoder can be further used in discriminative tasks. In this paper, we also make use of MIM to learn representations from different datasets. 

\subsection{Federated Leraning}
Although MIM can successfully learn to model a target domain without access to labeled examples, it still requires access to the unlabelled examples, and in the healthcare domain, this poses a privacy concern where in many cases, it is desired to keep the patients' data confidential. Moreover, since we have a diverse set of medical image classification tasks, we are going to deal with a lot of datasets, and moving all the data to the model will pose a huge communication overhead. Federated learning is a machine learning paradigm in which an algorithm is trained collaboratively via independent clients, each using their data. A centralized server is responsible for training an initial model and sending to clients upon request. Each client can train the model locally and only send model updates back to the server. The server can then aggregate updates from different clients and update its base model.

\section{Proposed Method}
\label{sec:method}
\subsection{Base Model}
The server utilizes a Vision transformer (ViT) as the backbone for the base model. The model is trained in a self-supervised way using masked image modeling. 

\subsection{Masked Image Modelling}
We utilize masked image modeling to pre-train the base model. Following the footsteps of \cite{he2022masked}, the input image is first divided into patches, the patches are split into two non-overlapping sets of visible, and hidden patches. The ViT-based encoder is fed with the visible patches, to which positional encoding is added as done with the standard vision transformer. The visible patches that the encoder works with are only around 25\% of the number of patches. This ensures that the model cannot leverage the redundancy present in most images and helps it learn the actual underlying distribution of the image. Finally, the encoder outputs a set of encoded visible patches which will be later processed by the decoder.

The decoder receives a set of encoded visible patches and mask tokens, where mask tokens are learnable and indicate the presence of a masked-out patch. Positional encoding is applied to all input tokens, and the decoder is tasked with reconstructing the pixel values of the masked patches.

\subsection{Stage-Wise Federated Learning}
After pre-training the base model via MIM, the server sends the base model $Z_b$ to a requesting client $C_i$ that aims to solve a medical image classification task with $N_i$ classes and $M_i$ training examples. The server appends a linear, fully connected layer to the base model and sends it to the client for fine-tuning. The client $C_i$ sends the model weights $Z_i$ back to the server. The server, after receiving $K$ client updates, declares a stage has ended and aggregates the received model updates via a weighted average with the formula:

\begin{equation}
    \acute{Z_b} = \frac{1}{\sum_{i}^{K}M_i}\sum_{i}^{1:K} Z_i \times M_i
\end{equation}

This allows the base model updates to not be dominated by clients who have few labeled examples. The updated model will be sent to the client in the next stages, will be the base for the fine-tuning done by the clients, and will eventually get sent back to the server. This effectively allows the server to keep a rolling update of its base parameters over time.

\subsection{Data Privacy}
Since each client reports back to the server the fine-tuned weights of the base model (as well as the number of training examples), the client maintains the privacy of their data points. Moreover, the client doesn't need to send the final linear classification layer either, further maintaining privacy by hiding the direct statistical correlation between the features and the final output.

\subsection{Model Trustworthiness}
To ensure that the server model doesn't diverge, a paired samples t-test will be conducted after every $S$ stage. The null hypothesis is that the means of the experiment and control populations are equal
$H_0: {\mu}_1 = {\mu}_2$. The alternative hypothesis (two-tailed) is that the two population means are not equal and is given by 
$H_1: {\mu}_1 \neq {\mu}_2$. The test statistic t is given by the formula:

\begin{equation}
    t = \frac{\bar{x}_d}{s_d / \sqrt{n}}
\end{equation}

Where $\bar{x}_d$ is the sample mean of the differences, $s_d$ is the sample standard deviation of the differences, and $n$ is the sample size. We also calculate the effect size for our t-test using Cohen's d measure. The d statistic is given by the formula:

\begin{equation}
    Cohen's D = \frac{{|\bar{x}_d|}}{s_d}
\end{equation}

The effect size calculation will be used to report if the difference is big enough since the results could be statistically significant but trivial. If the updated server model $\acute{Z_b}$ brings both significant and non-trivial improvements as compared to $Z_b$, it will be retained. If there is no improvement, the base model $Z_b$ will be retained to ensure the trustworthiness of our model performance across not only unseen test data but also across completely different tasks.

\section{Experiments and Results}
\label{sec:exp}
\subsection{Pre-training Datasets}
In order to achieve domain generalization, our pre-text pre-training task must span a diverse set of medical datasets.  Following the footsteps of \cite{anubhavthesis}, we utilize a diverse dataset that spans more than 10 different body parts as well as different image scans. Table \ref{tab:pretraindata} shows the datasets used for pre-training the server model.


\begin{table}[]
\resizebox{\columnwidth}{!}{\begin{tabular}{|l|l|c|}
\hline
\textbf{Body Part}             & \textbf{Dataset}         & \textbf{Scan Type}                    \\ \hline
\multirow{3}{*}{Chest}         & NIH Chest XRay \cite{wang2017chestx}           & X-Ray                                 \\ \cline{2-3} 
                               & RIDER Lung CT \cite{zhao2015data}, \cite{clark2013cancer}      & CT                                    \\ \cline{2-3} 
                               & LIDC-IDRI \cite{armato2015data}, \cite{clark2013cancer}            & CT / CR / DX                          \\ \cline{2-3} 
                                 \hline
\multirow{8}{*}{Lung}          & COVID-19-NY-SBU \cite{saltz2021stony},\cite{clark2013cancer}          & CR / CT / DX / MR / PT / NM / OT / SR \\ \cline{2-3} 
                               & COVID-19 CT \cite{an2020ct}, \cite{clark2013cancer}           & CT                                    \\ \cline{2-3} 
                               & MIDR-RICORD-1a \&1b \cite{tsai2020medical}, \cite{clark2013cancer}      & CT                                    \\ \cline{2-3} 
                               & QIN Lung CT  \cite{kalpathy2015qin}, \cite{clark2013cancer}            & CT                                    \\ \cline{2-3} 
                               & LungCT-Diagnosis \cite{grove2015quantitative}, \cite{clark2013cancer}        & CT                                    \\ \cline{2-3} 
                               & NSCLC-Radiomics-Genomics \cite{aerts2015data}, \cite{clark2013cancer} & CT                                    \\ \cline{2-3} 
                               & COVID-19-AR \cite{desai2020data}, \cite{clark2013cancer}             & CT / DX / CR                          \\ \cline{2-3} 
                               & CMB-LCA  \cite{CMB-LCA}, \cite{clark2013cancer}                & CT / DX / MR / NM / US                \\ \hline
Pancreas                       & Ctpred-Sunitinib-panNet \cite{CTpred-Sunitinib-panNET}, \cite{clark2013cancer}  & CT                                    \\ \hline
                               & Pancreas CT \cite{Pancreas-CT}, \cite{clark2013cancer}             & CT                                    \\ \hline
Colon                          & CT COLON ACRIN 6664 \cite{ACRIN6664}, \cite{clark2013cancer}      & CT                                    \\ \hline
                               & CMB-CRC   \cite{CMB-CRC}, \cite{clark2013cancer}              & CT / DX / MR / PT / US                \\ \hline
Bladder                        & TCGA-BLCA  \cite{TCGA-BLCA}, \cite{clark2013cancer}              & CT / CR / MR / PT / DX / Pathology    \\ \hline
Uterus                         & TCGA-UCEC   \cite{TCGA-UCEC}, \cite{clark2013cancer}             & CT / CR / MR / PT / Pathology         \\ \hline
\multirow{2}{*}{Kidney}        & TCGA-KIRC  \cite{TCGA-KIRC}, \cite{clark2013cancer}              & CT / MR / CR / Pathology              \\ \cline{2-3} 
                               & CPTAC-CCRCC  \cite{CPTAC-CCRCC}, \cite{clark2013cancer}            & CT / MR / Pathology                   \\ \hline
Prostate                       & CMB-PCA \cite{CMB-PCA}, \cite{clark2013cancer}                 & CT / MR / NM                          \\ \hline
Ovary                          & TCGA-OV \cite{TCGA-OV}, \cite{clark2013cancer}                 & CT / MR / Pathology                   \\ \hline
Rectum                         & TCGA-READ \cite{TCGA-READ}, \cite{clark2013cancer}               & CT / MR/ Pathology                    \\ \hline
\multirow{5}{*}{Miscellaneous} & Pseudo-PHI-DICOM-Data \cite{Pseudo-PHI-DICOM-Data}, \cite{clark2013cancer}   & CR / CT / DX / MG / MR / PT           \\ \cline{2-3} 
                               & Stagell-Colorectal-CT \cite{Stagell-Colorectal-CT}, \cite{clark2013cancer}    & CT                                    \\ \cline{2-3} 
                               & CT-ORG    \cite{CT-ORG}, \cite{clark2013cancer}               & CT                                    \\ \cline{2-3} 
                               & Pelvic-Reference-Data  \cite{pelvic-reference-data}, \cite{clark2013cancer}  & CT                                    \\ \cline{2-3} 
                               & TCGA-SARC   \cite{TCGA-SARC}, \cite{clark2013cancer}             & CT / MR / Pathology                   \\ \hline
\end{tabular}}
\caption{A breakdown by body parts of the datasets used to pre-train the server's base model. The dataset images are obtained using different scanning techniques.}

\label{tab:pretraindata}
\end{table}

\begin{figure}[!htb]

\begin{minipage}{.15\linewidth}
  \centering
  \centerline{\includegraphics[width=1.5cm]{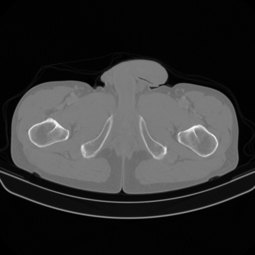}}
  \centerline{(a) Bladder}\medskip
\end{minipage}
\begin{minipage}{.15\linewidth}  \centering
  \centerline{\includegraphics[width=1.5cm]{images/nihchestxray_00000001_000.png}}
  \centerline{(b) Chest}\medskip
\end{minipage}
\begin{minipage}{.15\linewidth}
  \centering
  \centerline{\includegraphics[width=1.5cm]{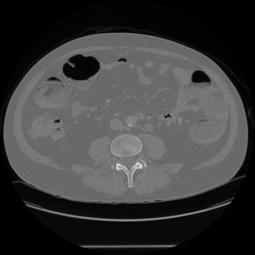}}
  \centerline{(c) Colon}\medskip
\end{minipage}
\begin{minipage}{.15\linewidth}
  \centering
  \centerline{\includegraphics[width=1.5cm]{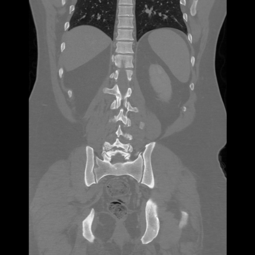}}
  \centerline{(d) Kidney}\medskip
\end{minipage}
\begin{minipage}{.15\linewidth}
  \centering
  \centerline{\includegraphics[width=1.5cm]{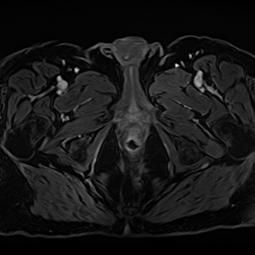}}
  \centerline{(e) Lung}\medskip
\end{minipage}
\begin{minipage}{0.15\linewidth}
  \centering
  \centerline{\includegraphics[width=1.5cm]{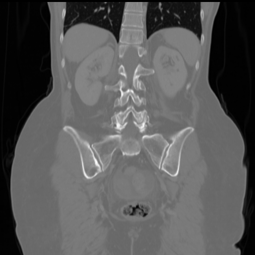}}
  \centerline{(f) Ovary}\medskip
\end{minipage}
\begin{minipage}{.15\linewidth}
  \centering
  \centerline{\includegraphics[width=1.5cm]{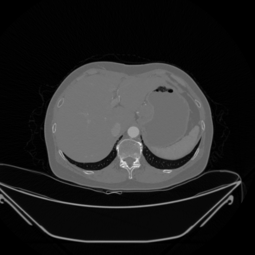}}
  \centerline{(g) Pancreas}\medskip
\end{minipage}
\begin{minipage}{.15\linewidth}
  \centering
  \centerline{\includegraphics[width=1.5cm]{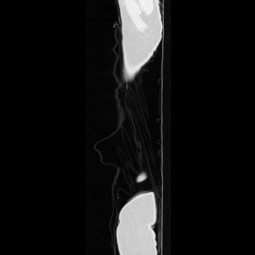}}
  \centerline{(h) Prostate}\medskip
\end{minipage}
\begin{minipage}{0.15\linewidth}
  \centering
  \centerline{\includegraphics[width=1.5cm]{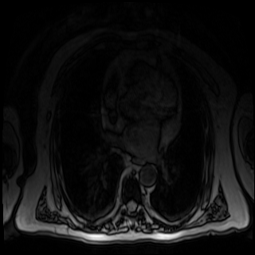}}
  \centerline{(i) Rectum}\medskip
\end{minipage}
\begin{minipage}{0.15\linewidth}
  \centering
  \centerline{\includegraphics[width=1.5cm]{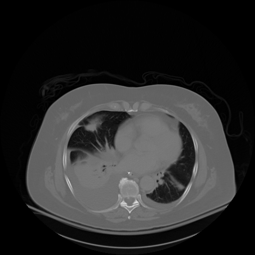}}
  \centerline{(j) Uterus}\medskip
\end{minipage}
\begin{minipage}{\linewidth}
  \centering
  \centerline{\includegraphics[width=0cm]{images/TCGA-UCEC0000fcfc-bc5d-422c-b521-445a689d9024_1-005.dcm.png}}
\end{minipage}
\caption{Samples of the pretraining dataset.}
\label{fig:pretrain}
\end{figure}

\subsection{Stage-wise Federated Learning}

We collected $8$ additional public datasets that span different medical imaging applications, as well as a huge private dataset. The datasets span different medical applications.

We split the 8 datasets into 4 stages, with 2 clients per stage to be able to test our approach on as many stages as possible. Each stage represents clients that train concurrently. After the training ends, the server gets updated as mentioned in the methodology. To test the efficacy of our model, starting from stage 2, we will repeat all the experiments twice. Once starting from the pre-trained weights of the server (the control group) and once from our updated server model (the test group). The results are presented in table \ref{tab:expdata}

\textbf{Stage 1 datasets} comprise a 1) \textit{brain tumor classification dataset} \cite{braintumormridata} which has 3264 MRI images. The task is to classify images into three types of different tumors as well as a no-tumor class. 2) \textit{White blood cells} 5-class classification dataset \cite{kouzehkanan2022large}. white blood cells, also known as leukocytes, consist of phagocytes and lymphocytes. Phagocytes act quickly upon infection, while lymphocytes play a role in the acquired immune response. Phagocytes can be further categorized into monocytes, eosinophils, basophils, and neutrophils.

\textbf{Stage 2 datasets} comprise a 1) \textit{Breast Histopathology Images dataset} \cite{cruz2014automatic}, \cite{janowczyk2016deep}. The dataset benchmarks the classification of Invasive Ductal Carcinoma (IDC), a common subtype of breast cancer. The dataset comprises 198,738 IDC(-) and 78,786 IDC(+) histopathology image patches. 2) \textit{NCT-CRC-HE-100K} \cite{kather2018100} is a dataset comprising 100,000 image patches from H\&E stained histological images of human colorectal cancer (CRC) and normal tissue. Each patch is 224x224 pixels with a resolution of 0.5 microns per pixel. The dataset includes nine tissue classes, color-normalized using Macenko's method \cite{macenko2009method}. 

\textbf{Stage 3 datasets} comprise a 1) \textit{COVID-19 Radiography Dataset} \cite{chowdhury2020can}, \cite{rahman2021exploring} is a big collection of X-Ray images which were built on multiple different datasets. This collection comprises COVID-positive (3615), normal (10192 images), Lung Opacity (6012 images), and viral Pneumonia (1345 images). 2) \textit{Breast Ultrasound Images Dataset} \cite{al2020dataset} comprises breast ultrasound images from women aged 25 to 75, collected in 2018, involving 600 female patients. The dataset contains 780 PNG format images. The images are classified into three categories: normal, benign, and malignant.

\textbf{Stage 4 datasets} comprise a 1)\textit{Ocular Toxoplasmosis Fundus Images} dataset \cite{cardozo2023dataset} comprises 822 images representing individuals with suspected congenital toxoplasmosis infection. The fundus images are categorized into two main groups, Healthy and Diseased, presenting a binary classification task. 2) \textit{BC Breast Cancer} dataset, which is a private dataset that belongs to the BC Cancer Agency breast cancer chest CT dataset. There are a total of 29,686 patients in this dataset. The dataset contains 57,807 CT series for the patients, taken through the course of the patient’s treatment for breast cancer. Each series contain a range of dice slices, 100 slices per series making it more than 5 million patient scans. The data cohort was collected from the year 1980-2017
\subsection{Model Trustworthiness: Paired T-test}
We performed a paired sample T-test with a two-tailed distribution (degrees of freedom: 5) after the fourth stage to evaluate the statistical significance of our results compared to the control group. The test was conducted using the common significance level of $\alpha = 0.05$. The paired t-test results indicated a significant and substantial difference between the control group (mean = 89.9, standard deviation = 7.8) and the experimental group (mean = 92.5, standard deviation = 6.9), yielding $t(5) = 2.7$ and $\mathit{p-value} = .044$. As the $\mathit{p-value}$ is less than $\alpha$, we reject the null hypothesis, indicating that the population average of the experimental group (i.e., the group undergoing fine-tuning with the stage-wise federated approach) is not equal to that of the control group. The observed effect size, represented by d, is large at 1.09, signifying a substantial difference between the average of the observed differences and the expected average of the differences.


\begin{table}[]
\resizebox{\columnwidth}{!}{
{\renewcommand{\arraystretch}{2}
\begin{tabular}{|c|c|c|c|c|}
\hline
Stage                    & Dataset                        & Epochs & Control Accuracy & Experiment Accuracy    \\ \hline
\multirow{2}{*}{Stage 1} & brain-tumor-classification-mri & 10     & 80.71\%          & -                \\ \cline{2-5} 
                         & white-blood-cells & 10     & 98.09\%          & -                \\ \hline
\multirow{2}{*}{Stage 2} & IDC                            & 10     & 87.66\%          & \textbf{88.2\%}  \\ \cline{2-5} 
                         & NCT-CRC-HE-100k                & 15     & 96.18\%          & \textbf{96.57\%} \\ \hline
\multirow{2}{*}{Stage 3} & COVID-19\_Radiography\_Dataset & 50     & 93.41\%          & \textbf{94.23\%} \\ \cline{2-5} 
                         & Breast-Ultrasound              & 100    & 75.15\%          & \textbf{80.89\%} \\ \hline
\multirow{2}{*}{Stage 4} & ocular-toxoplasmosis-fundus    & 100    & 95.18\%          & \textbf{100\%}   \\ \cline{2-5}
                         & BC Breast Cancer               & 80     & 92\%             & \textbf{95\%}    \\ \hline
\end{tabular}}}
\caption{The table exhibits the classification accuracies of clients across eight distinct tasks. Clients in the later stages experience greater advantages from the accumulated knowledge accessible at the server, thereby demonstrating a more pronounced performance gap between the experimental and control groups.}
\label{tab:expdata}
\end{table}

\begin{figure}
\centering
    \includegraphics[width=\linewidth]{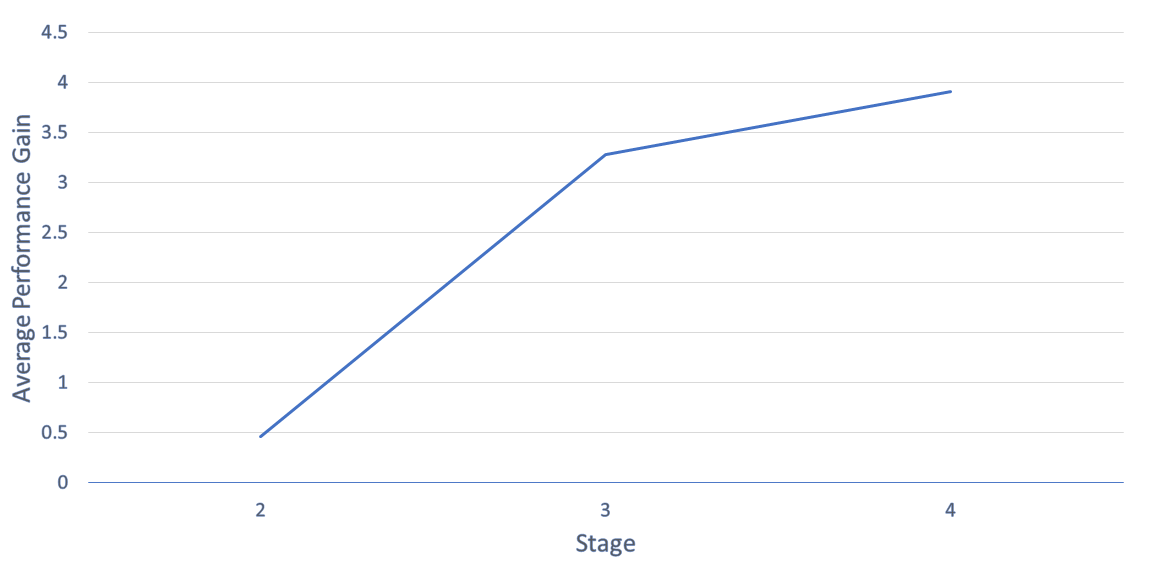}
    \caption{The plot shows the average accuracy gain across the different stages. As more clients send their fine-tuned weights back to the server, the server can aggregate more knowledge and thus present a more robust model to clients in the next stages.}
    \label{fig:fed}
\end{figure}
\label{fig:stage_improve}


\section{Conclusion}
We presented a novel approach to generalize across domains of various medical imaging applications. We showed that our federated approach maintains a model that increases in robustness over time when new clients utilize and finetune the base server model and then send the fine-tuned weights back to the server. While clients in the different stages of our experiment had local data from significantly different domains, they were still able to benefit from the aggregated knowledge from the previous stages.  

\section{Acknowledgment}
The authors acknowledge Isha Shah for validating the approach on the large-scale private BC cancer dataset. The authors also acknowledge 1) the National Cancer Institute and the Foundation for the National Institutes of Health and 2) The Multi-national NIH Consortium for CT AI in COVID-19. 3)the National Cancer Institute's Cancer Moonshot Biobank, 4) the TCGA Research Network, 5) the National Cancer Institute Clinical Proteomic Tumor Analysis Consortium (CPTAC) for providing the data which was used as a part of our pre-training dataset

\bibliographystyle{IEEEbib}
\bibliography{refs}

\end{document}